\documentclass[
superscriptaddress,
preprint,
amsmath,
amssymb,
aps,
pra,
]{revtex4-2}
\usepackage{graphicx}
\usepackage{hyperref}

\begin{document}

\title{A revisit on the hydrogen atom induced by a uniform static electric field}
\author{Tran Duong Anh-Tai}
\affiliation{Department of Physics, Ho Chi Minh City University of Education, Ho Chi Minh City, Vietnam}
\affiliation{Quantum Systems Unit, OIST Graduate University, Onna, Okinawa 904-0495, Japan}
\author{Le Minh Khang}
\affiliation{Department of Physics, Ho Chi Minh City University of Education, Ho Chi Minh City, Vietnam}

\author{Nguyen Duy Vy}
\affiliation{Laboratory of Applied Physics, Science and Technology Advanced Institute, Van Lang University, Ho Chi Minh City, Vietnam\looseness=-1}
\affiliation{Faculty of Applied Technology, School of Technology, Van Lang University, Ho Chi Minh City, Vietnam\looseness=-1}\email{nguyenduyvy@vlu.edu.vn}

\author{Thu D.H. Truong}
\affiliation{Department of Physics, Ho Chi Minh City University of Education, Ho Chi Minh City, Vietnam}

\author{Vinh N.T. Pham}
\affiliation{Department of Physics, Ho Chi Minh City University of Education, Ho Chi Minh City, Vietnam}
\affiliation{Postgraduate Studies Office, Ho Chi Minh City University of Education, Ho Chi Minh City, Vietnam\looseness=-1}
\email[Corresponding author: ]{vinhpnt@hcmue.edu.vn}

\date{\today{}}

\begin{abstract}
In this paper, we revisit the Stark effect of the hydrogen atom induced by a uniform static electric field. In particular, a general formula for the integral of associated Laguerre polynomials was derived by applying the method for Hermite polynomials of degree n proposed in the work [Anh-Tai T.D. et al., 2021 AIP Advances \textbf{11} 085310]. The quadratic Stark effect is obtained by applying this formula and the time-independent non-degenerate perturbation theory to hydrogen. Using the Siegert State method, numerical calculations are performed and serve as data for benchmarking. The comparisons are then illustrated for the ground and some highly excited states to provide an insightful look at the applicable limit and precision of the quadratic Stark effect formula for other atoms with comparable properties. 
\end{abstract}


\maketitle

\tableofcontents{}

\section{Introduction}
\label{intro}
Atomic/molecular ionization is one of the most fundamental quantum mechanical phenomena, and a comprehensive understanding of its mechanism could have implications for a variety of physics topics. In the last three decades, strong electric field ionization of atoms and molecules has attracted a great deal of experimental and theoretical interest due to its abundance of nonlinear physical processes, such as high-harmonic generation (HHG) \cite{chelkowski1990efficient,lewenstein1994theory,winterfeldt2008colloquium} and nonsequential double ionization (NSDI) \cite{suran1975observation,rudenko2007correlated,bergues2012attosecond,staudte2007binary,truong2022soft,truong2022intensity}. In the presence of a weak electric field, the spectral levels of the atomic bound states are typically split and shifted, resulting in the well-known Stark effect \cite{stark1913observation,stark1914beobachtungen1,stark1914beobachtungen2}. Epstein introduced the first theoretical explanation of this effect using old quantum mechanics \cite{epstein1916theorie}. Subsequently, Schr\"{o}dinger also provided an explanation for the hydrogen Stark effect as a part of the formalism of wave mechanics \cite{schr1926quantisierung}. Epstein then independently revisited the problem using wave mechanics \cite{epstein1926}. It is noteworthy to note that Epstein solved the problem using parabolic coordinates, which have more advantages than spherical coordinates and are widely employed in the study of strong-field ionization of atoms and molecules \cite{batishchev2010atomic,pham2014,ohgoda2017photoionization,trinh2015weak,pham2019images}. 

Various approaches, including the WKB \cite{wentzel1926,bekenstein1969}, group theory \cite{alliluev1974}, hypervirial relations \cite{lai1981}, numerical methods \cite{damburg1976,damburg1978,damburg1979,damburg1981}, and Bender-Wu asymptotic formula \cite{benassi1979} have been used to study this effect. A generalization to arbitrarily high-order perturbed corrections \cite{silverstone1978,alliluev1980,bolgova2003}, have also been revisited. Intriguingly, experimental imaging of the nodal structure of hydrogen Stark states has recently revealed a significant finding \cite{stodolna2013}. Consequently, it is evident that the Stark effect and the hydrogen atom continue to play vital roles in modern physics.

The Rayleigh-Schr\"{o}dinger perturbation theory is a fundamental and essential approximation method in quantum mechanics and is a typical topic in standard quantum mechanics textbooks \cite{landau1958,bethe1957,magnasco2006elementary,schwabl2007quantum,shankar2012principles,razavy2013quantum,fernandez2013,griffiths2018introduction}. Since the calculation is simple enough, the derivation of the Stark effect of the hydrogen atom with the second-order correction, also known as the quadratic Stark effect, would be an outstanding illustration for perturbation theory at the undergraduate level. 

In order for the perturbation theory to be legitimate, the applied electric field should be sufficiently small, resulting in a weak electric-field strength. However, defining a weak electric field is not trivial. In early experiments \cite{rottke1986photoionization,glab1985spectroscopy,glab1985stark}, field strengths of $10^{-7}$--$10^{-6}$ a.u. (~10$^3$--10$^4$ V/cm) were considered strong; however, in more recent experiments \cite{stodolna2013,cohen2013wave,cohen2016photoionization}, a field strength of $0.1$ a.u. is regarded as weak. To our knowledge, the applicable interval for the approximation of the formula characterizing the Stark effect derived from perturbation theory has not yet been thoroughly investigated. Consequently, it is essential to revisit this effect of hydrogen atom and also consider the relating issues that are still open to discuss.

Our objective is to present a mathematically rigorous and straightforward method for approximating the eigenvalues of a hydrogen atom induced by a uniform static electric field as an illustration of the use of advanced one-variable calculus to solve fundamental quantum mechanics problems.  More specifically, quantum mechanics is related to perturbation theory and the non-relativistic hydrogen atom in parabolic coordinates, while advanced one-variable calculus is represented by the integral of associated Laguerre polynomials. In our approach, the operators are expressed in parabolic coordinates, which are efficient and widely used for atomic ionization studies. Using the parabolic coordinates effectively reduces the number of analytical calculations, as the integrals of the associated Laguerre polynomials can be calculated using the general formula. Moreover, we numerically solve the Schr\"{o}dinger equation of the system by employing the Siegert State (SS) method \cite{batishchev2010atomic,pham2014,ohgoda2017photoionization} (see Appendix B for more details) in order to evaluate the exactness and the range of field strength such that the approximated formulas are still applicable. A comparison is illustrated for the ground state and some highly excited states with the principal quantum number up to $n = 10$.

The structure of the article is as follows. In Sec. \ref{sec_analytical}, we present the derivation of the formula characterizing the quadratic Stark effect of hydrogen as an application of using advanced one-variable calculus to solve quantum mechanical problems, as well as comparisons of the approximate formula and numerical calculations. The conclusion is presented in Sec. III. This article uses the atomic units in which $\hbar = m_e = e = 1$ for the sake of simplification. $X^{(n)}$ denotes the $n$th-order perturbative correction to the physical quantity $X$. Appendix A reformulates the integral of Laguerre polynomials, while an overview of the Siegert state method is presented in Appendix B.

\section{Results and discussion} \label{sec_analytical}
\begin{figure}[tb] 	\centering
\includegraphics[width=0.75\textwidth]{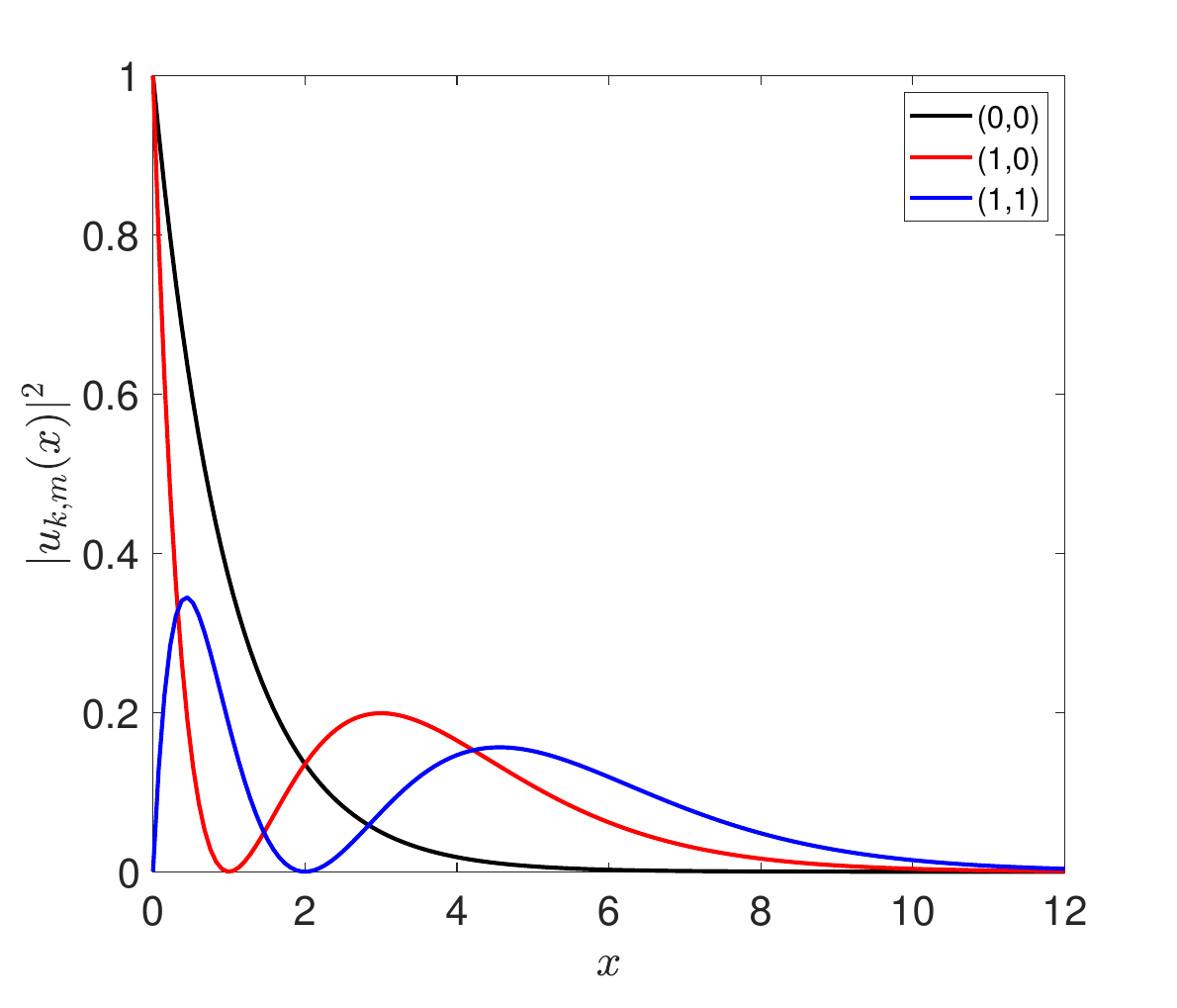}
\caption{The absolute squared of $u_{k,m}(x)$ for three sets of $(k,m)$. Note that in the plot  $\int\limits_0^\infty |u_{k,m}(x)|^2dx = 1$.}
	\label{fig:u_km}
\end{figure}

For the eigenvalues of the non-relativistic hydrogen induced by a uniform z-axis directed static electric field, we start from the Hamiltonian
\begin{align}
	\label{eq1} 
	\widehat{H} = -\dfrac{1}{2}\Delta - \frac{1}{r} + Fz,
\end{align}	
where the first two terms are usually written as 
\begin{align}
	\label{coulomb} 
	\widehat{H}_0 = -\dfrac{1}{2}\Delta - \frac{1}{r},
\end{align}	
which characterizes the non-relativistic hydrogen atom in the Coulomb field only, and $Fz$ is the potential resulting from the uniform static electric field with the strength $F \geq 0$. Here, $r$ and $z$ represent the distance and $z$-axis displacement of the electron from the proton, which is assumed to be at the origin of the coordinate system, respectively. Instead of using the spherical coordinates, in this work we employ the parabolic coordinates which are orthogonal and are formulated by
\begin{equation}
	\label{eq2} 
	\xi =r+z\left( 0\le \xi < +\infty  \right),\eta =r-z\left( 0\le \eta < +\infty  \right),\varphi \left( 0\le \varphi \le 2\pi\right),
\end{equation}	
where ($\xi, \eta, \varphi$) corresponds to the $\xi$ distance, the $\eta$ distance, and the azimuthal angle $\varphi$, respectively. The infinitesimal element of volume is then given by the expression
\begin{equation}
	\label{jacobian}
	dV = \dfrac{\xi+\eta}{4}d\xi d\eta d\varphi, 
\end{equation}
while the Laplace operator, $\Delta$, is given as follows  
\begin{equation}
	\Delta = \dfrac{4}{\xi+\eta} \dfrac{\partial}{\partial \xi} \left(\xi \dfrac{\partial}{\partial \xi}\right) + \dfrac{4}{\xi+\eta} \dfrac{\partial}{\partial \eta} \left(\eta \dfrac{\partial}{\partial \eta}\right) + \dfrac{4}{\xi\eta}\dfrac{\partial^2}{\partial\varphi^2}.
\end{equation}
In the parabolic coordinates, the Schr\"{o}dinger equation (SE) describing the non-relativistic hydrogen in the presence of the static electric field is explicitly given by
\begin{equation}
    \label{SE}
    \left[\dfrac{\partial}{\partial \xi} \left(\xi \dfrac{\partial}{\partial \xi}\right) +  \dfrac{\partial}{\partial \eta} \left(\eta \dfrac{\partial}{\partial \eta}\right) + \dfrac{1}{4}\left(\dfrac{1}{\xi} + \dfrac{1}{\eta}\right)\dfrac{\partial^2}{\partial\varphi^2} + \dfrac{1}{2}E\left(\xi+\eta\right) + 1 -\dfrac{F}{4}\left(\xi^2 - \eta^2\right) \right]\psi(\xi,\eta,\varphi) = 0.
\end{equation}
In the absence of the electric field $F=0$, Eq. \eqref{SE} have analytical solutions associating in the bound states of the non-relativistic hydrogen atom in the pure Coulomb field in the parabolic coordinates that have been rigorously presented in Refs. \cite{bethe1957,landau1958,tai2015on}. In the following we briefly present the necessary results for the derivation of quadratic Stark effect. The parabolic coordinates enable us to express the bound-state eigenfunctions of $\widehat{H}_0$ as 
\begin{equation}
	\label{eq3} 
	\psi^{(0)}_{n_1,n_2,m}(\xi,\eta,\varphi) = Au_{n_1,m}\left(\xi/n\right)u_{n_2,m}\left(\eta/n\right)\exp(im\varphi),
\end{equation}	
where
\begin{equation}
	\label{eq6} 
	A=\dfrac{1}{n^2}\sqrt{\dfrac{n_1!n_2!}{\pi(n_1+m)!(n_2+m)!}}
\end{equation}
is the normalization factor. Meanwhile, the bound-state eigenvalues of $\widehat{H}_0$  are given as 
\begin{equation}
	\label{E0}
	E^{(0)} = -\dfrac{1}{2n^2}, 
\end{equation}	
where $n$ is the principal quantum number. Hence, a bound state is denoted by a set of three quantum numbers $(n_1, n_2, m)$, where $n_1$ and $n_2$ represent the parabolic quantum numbers describing the $\xi$ and $\eta$ variables, respectively, and $m$ is the magnetic quantum number. Note that in the calculations that follow, the magnetic quantum number should be positive because the eigenfunctions $\psi^{(0)}_{n_1,n_2,m}(\xi,\eta,\varphi)$ with $\pm m$ are degenerate, as is evident from Eq. \eqref{eq3}. The relationship between the three quantum numbers $(n_1,n_2,m)$ and the principal quantum number $n$ is
\begin{equation}
	n = n_1 + n_2 + |m| + 1.
\end{equation}
The functions $u_{n_1,m}(\xi/n)$ and $u_{n_2,m}(\eta/n)$,  which independently describe the dependence of eigenfunctions along $\xi$ and $\eta$ coordinates, have the identical form given by
\begin{equation}
	\label{eq8}                                            
	u_{k,m}\left(x\right) = x^{m/2}\exp\left(-x/2\right)L_{k}^{m}\left(x\right),         
\end{equation}	
where $L_{k}^{m}(x)$ is the associated Laguerre polynomials. Physically, the parabolic quantum number $k=\{n_1,n_2\}$ determines the number of $u_{k,m}\left(x\right)$ nodes, whereas the magnetic quantum number $m$ is associated with the boundary condition at $x = 0$. For $m = 0$, $u_{k,0}\left(0\right)$ decays from 1, whereas $u_{k,m\neq 0}\left(0\right) = 0$ for $m \ne 0$. This characteristic is evident in Fig. \ref{fig:u_km}, which depicts the absolute squared of $ u_{k,m}\left(x\right)$ for various values $(k,m)$. 

In the presence of the electric field, from Eq. \eqref{eq1}, eigenvalues cannot be determined analytically. Consequently, the time-independent non-degenerate perturbation theory is utilized to approximate the required eigenvalues. In this method, it is assumed that the uniform static electric field is sufficiently weak, so that the potential caused by such a field can be regarded as a small perturbation. In the present study, the eigenvalues of the Hamiltonian \eqref{eq1} are expanded up to the second order of the electric field strength
\begin{equation}
	\label{expansion}
	E = E^{(0)} + E^{(1)} F +  E^{(2)} F^2 + \mathcal{O}(F^3).
\end{equation}
Here $E^{(0)}$, $E^{(1)}$ and $ E^{(2)}$ are the zero-, first- and second-order corrections to the eigenvalues, respectively. $E^{(0)}$ coincides to the eigenvalues of hydrogen atom in the Coulomb field given by Eq. \eqref{E0}. $E^{(1)}$ and $ E^{(2)}$ are then determined by the time-independent non-degenerate perturbation theory. $E^{(1)}$ is obtained by evaluating
\begin{align}     \label{firstcorr}
	E^{(1)} & = \langle \psi^{(0)}_{n_1,n_2,m}|  \dfrac{\xi-\eta}{2} | \psi^{(0)}_{n_1,n_2,m} \rangle \nonumber \\ 
	& = 2\pi A^2 \int\limits_0^\infty \int\limits_0^\infty u^*_{n_1,m}\left(\xi/n\right)u^*_{n_2,m}\left(\eta/n\right)\dfrac{\xi-\eta}{2} u_{n_1,m}\left(\xi/n\right)u_{n_2,m}\left(\eta/n\right) \dfrac{\xi+\eta}{4}d\xi d\eta,
\end{align}	
where $u^*_{k,m}\left(x\right) = u_{k,m}\left(x\right)$, the complex conjugate of $u_{k,m}\left(x\right)$. Note that in the following, the results of the integrations over $\xi$ and $\eta$ will be automatically multiplied by 2$\pi$, the factor coming from the integration over the azimuthal variable, $\varphi$.
Hence, Eq. \eqref{firstcorr} can be rewritten as
\begin{equation}
	\label{firstcorr_subtract}
	E^{(1)} = E_{1}^{(1)} - E_{2}^{(1)}
\end{equation}
where
\begin{align}
	\label{eq11} 
	& E_{1}^{(1)} =\dfrac{\pi n^2 A^2}{4} \int\limits_{0}^{\infty}\left(\xi/n\right)^{2}u^2_{n_1,m}\left(\xi/n \right)d\left(\xi/n\right)\int\limits_{0}^{\infty}u^2_{n_2,m}\left(\eta/n \right)d\left(\eta/n\right),\\
	\label{eq12} 
	& E_{2}^{(1)}=\dfrac{\pi n^2 A^2}{4} \int\limits_{0}^{\infty}\left(\eta/n\right)^{2}u^2_{n_2,m}\left(\eta/n \right)d\left(\eta/n\right)\int\limits_{0}^{\infty}{u^2_{n_1,m}}\left(\xi/n \right)d\left(\xi/n\right).           
\end{align}	
Obviously, the integrals  $E_{1}^{(1)}$, and $E_{2}^{(1)}$ can be rewritten in an unified form as
\begin{align}
	\label{unified} 
	I_{k,\ell} =\dfrac{\pi n^2 A^2}{4}\int\limits_{0}^{\infty} x^2 u^2_{k,m}(x)dx\int\limits_{0}^{\infty}u^2_{\ell,m}(y)dy.
\end{align}	
The integral $I_{k,\ell}$ can be evaluated by using the Eq. \eqref{laguerre_int}, hence we get
\begin{align}
	\label{E11} 
	E_{1}^{(1)} = I_{n_1,n_2} = \dfrac{1}{4}\left(6n_1^2+6n_1m+6n_1+m^2+3m+2\right), \\
	\label{E21} 
	E_{2}^{(1)} = I_{n_2,n_1} = \dfrac{1}{4}\left(6n_2^2+6n_2m+6n_2+m^2+3m+2\right).         
\end{align}	
Plugging Eqs. \eqref{E11} and \eqref{E21} to Eq. \eqref{firstcorr_subtract}, we obtain	
\begin{equation}
	\label{first_correction} 
	E^{(1)} = \dfrac{3}{2}n(n_1-n_2). 
\end{equation}	
With this first-order correction, the eigenvalue in an electric field is given as
\begin{equation}
	\label{linear_stark}
	E = -\dfrac{1}{2n^2} + \dfrac{3}{2}n(n_1-n_2) F + \mathcal{O}(F^2). 
\end{equation}
Equation \eqref{linear_stark} presents the linear Stark effect since the splitting of the energy levels is
\begin{equation}
	\Delta E^{(1)} = \dfrac{3}{2}n(n_1-n_2) F \propto F.
\end{equation}
Moreover, substates with identical parabolic quantum numbers, $n_1 = n_2$, do not perceive the applied electric field with the first-order field-induced correction because their distribution of the electrical charge between the $\xi$ and $\eta$ coordinates is symmetric, whereas the linear Stark effect derives from the asymmetric distribution of the electrical charge \cite{bethe1957,landau1958}.

The second-order correction to the eigenvalues can be computed as follows
\begin{equation}
	\label{secondcorr} 
	E^{(2)} = \langle \psi^{(0)}_{n_1,n_2,m} | \dfrac{\xi-\eta}{2}|\psi^{(1)}_{n_1,n_2,m}\rangle 
\end{equation}	
where $|\psi^{(1)}_{n_1,n_2,m}\rangle$ is the first-order correction to the eigenfunctions which has been derived by using the Green's function approach in Ref. \cite{kamenski2000} and reads
\begin{equation}
	\label{firstcorr_wavefunction}
	\psi^{(1)}_{n_1,n_2,m}(\xi,\eta,\varphi) = \dfrac{n^3A}{4}\left[\Phi_{n_1,m}\left(\xi/n\right)u_{n_2,m}\left(\eta/n\right)-u_{n_1,m}\left( \xi/n\right)\Phi_{n_2,m}\left(\eta/n\right)\right]\exp(im\varphi),
\end{equation}	
in which
\begin{multline}
	\label{Phi} 
	\Phi_{k,m}(x)=\dfrac{(k+m)(k+m-1)}{2}u_{k-2,m}(x)-(3n-m-3-2n_1)(k+m)u_{k-1,m}(x) +\\+3ku_{k,m}(x)-(3n-m+1-2n_1)(k+1)u_{k+1,m}(x)+\dfrac{(k+1)(k+2)}{2}u_{k+2,m}(x).  
\end{multline}
We note that Eq. \eqref{firstcorr_wavefunction} is very important for the derivation of the second-order correction to the eigenvalues in the following, but the details of which are not trivial. Therefore, we reuse a part of the results presented in Ref. \cite{kamenski2000}.
Combining Eqs. \eqref{eq3}, \eqref{Phi}, and \eqref{firstcorr_wavefunction} again yields 
\begin{align}
	\label{eq19} 
	E^{(2)} = E_{1}^{(2)}+E_{2}^{(2)},
\end{align}
where the two integrals $E_{1}^{(2)}$ and $E_{2}^{(2)}$ have an unified form as 
\begin{multline}
	\label{eq20}
	K_{i,j} = \dfrac{\pi n^7A^2}{16}\left\{\int\limits_{0}^{\infty} x^2 \Phi_{i,m}\left(x\right) u_{i,m}\left(x\right)dx \int\limits_{0}^{\infty}u_{j,m}\left(y\right) u_{j,m}\left(y\right) dy \right. + \\ \left.
	-\int\limits_{0}^{\infty} x^2 u_{i,m}\left(x\right)u_{i,m}\left(x\right) dx \int\limits_{0}^{\infty} \Phi_{j,m}\left(y\right) u_{j,m}\left(y\right) dy \right\}.
\end{multline}	
Again, the integral $K_{i,j}$ can be evaluated and derived with the help of Eq.  \eqref{laguerre_int}
\begin{align}
	\label{kij} 
	K_{i,j} = & \dfrac{n^3}{16}\left[ 4i(i^2-1) +6i(i+im+m^2+m-1) \right. + \nonumber \\ 
	& \left.+ (m^2+3m+2)(3i-3j+2m-3) +18(i^2+im+i)(i-j-2n)\right].                        
\end{align}
Substituting $E^{(2)}_1 = K_{n_1,n_2}$ and $E^{(2)}_2 = K_{n_2,n_1}$ into Eq. \eqref{eq19} yields	
\begin{equation}
	\label{secondcorr_eigvals} 
	E^{(2)} = -\dfrac{n^4}{16}\left[17n^2-3(n_1-n_2)^2-9m^2+19\right].
\end{equation}
Consequently, the formula that describes the so-called quadratic Stark effect is obtained
\begin{equation}
	\label{quadratic_stark}
	E = -\dfrac{1}{2n^2} + \dfrac{3n(n_1-n_2)}{2} F - \dfrac{n^4}{16}\left[17n^2-3(n_1-n_2)^2-9m^2+19\right]F^2 + \mathcal{O}(F^3). 
\end{equation}	
In contrast to the linear Stark effect, the quadratic Stark effect involves the splitting of energy levels
\begin{equation}
	\Delta E^{(2)} = \dfrac{3n(n_1-n_2)}{2} F - \dfrac{n^4}{16}\left[17n^2-3(n_1-n_2)^2-9m^2+19\right]F^2,
\end{equation}
exists even when the electrical charge distribution is symmetric, $n_1=n_2$. This indicates that the substates are always able to detect the presence of the electric field, despite of its weakness. When the atom is exposed to a uniform static electric field, the degeneracy of substates with the same principal quantum number $n$ is completely eliminated \cite{landau1958,bethe1957}.

\begin{figure}[htb!] 	\centering 
\includegraphics[width=1.\textwidth]{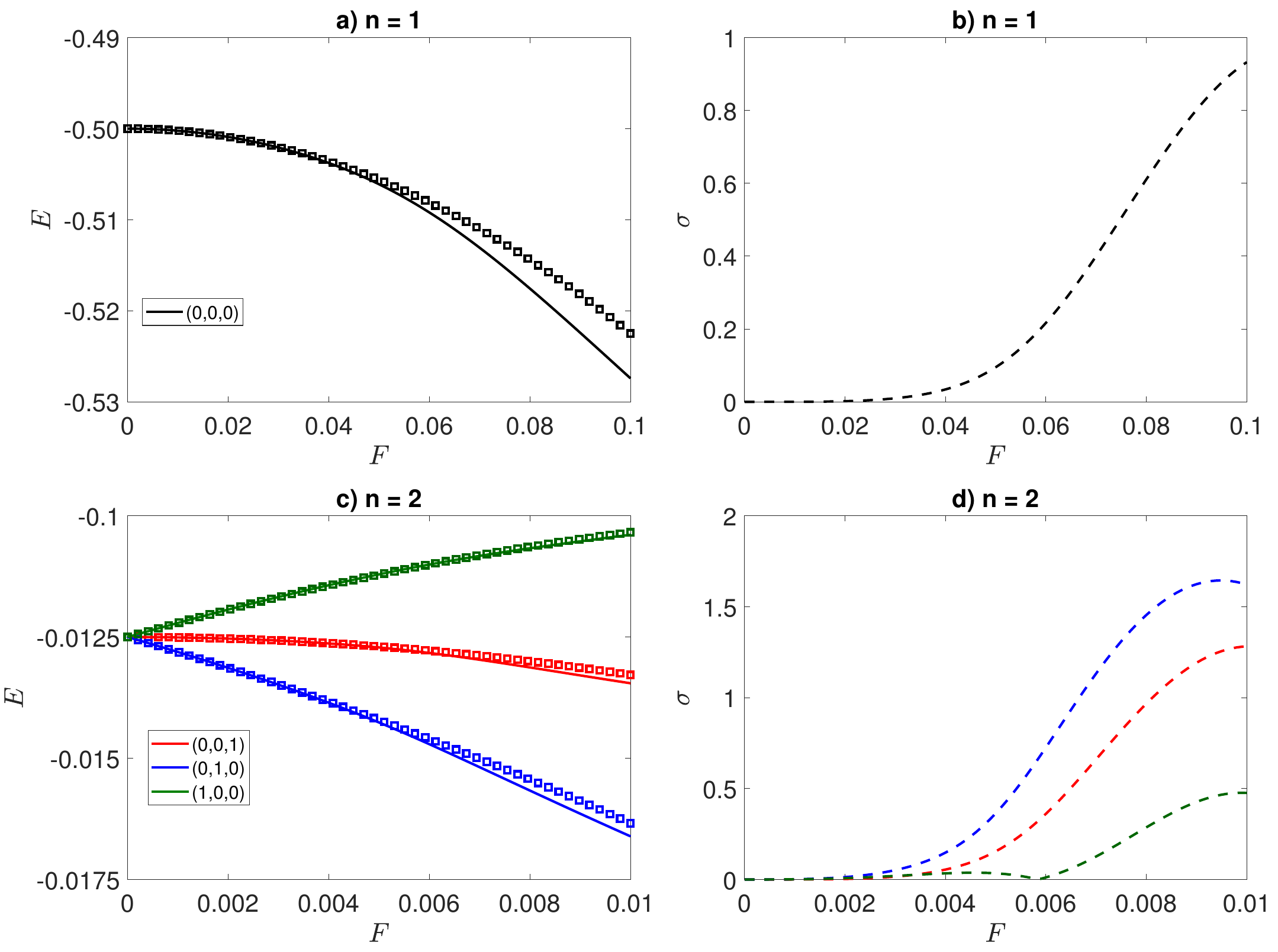}
	\caption{The first column illustrates the dependence of the eigenvalue, $E$, on the external field strength, $F$ for the ground state and the first excited state of the hydrogen atom induced by a static electric field. In the second column, the relative errors between the numerical result obtained by the Siegert State method and the approximated formula are displayed. Note that the solid line represents the numerical results, whereas the squares represent the approximations.} 
	\label{fig:gs_excited}
\end{figure}

Using the Siegert State (SS) method, we explicitly verify the exactness and applicability of the approximated formula given by the equation \eqref{quadratic_stark} using numerical calculations. Thus, we compute the relative deviation specified by 
\begin{equation}
	\sigma = \left|\dfrac{E_{ana}-E_{num}}{E_{num}}\right|\times 100\%,
\end{equation}
where $E_{num}$ and $E_{ana}$ represent the results of numerical and analytical methods, respectively. 
Figure  \ref{fig:gs_excited} illustrates explicitly the evaluation for the ground and the first excited states with three substates of hydrogen induced by a uniform static electric field. For the ground state, it can be observed that in the regime $0 \leq F \leq 0.5$ a.u., the results obtained by the approximated formula and numerical calculation are well-matched, and the relative error is close to $0$. The discrepancy between two approaches increases as the electric-field strength increases; however, it is still less than $1\%$ at $F=0.1$ a.u. 
In Figs. \ref{fig:gs_excited}(c)--(d), the three substates of the first excited state exhibit the same characteristic. When $F=0.01$ a.u., the $(0,1,0)$ substate has the maximum relative deviation, below $2\%$, while the $(1,0,0)$ substate has the smallest relative error, approximately $0.5\%$.

\begin{figure}[tb] 
	\centering 
	\includegraphics[width=1.05\textwidth]{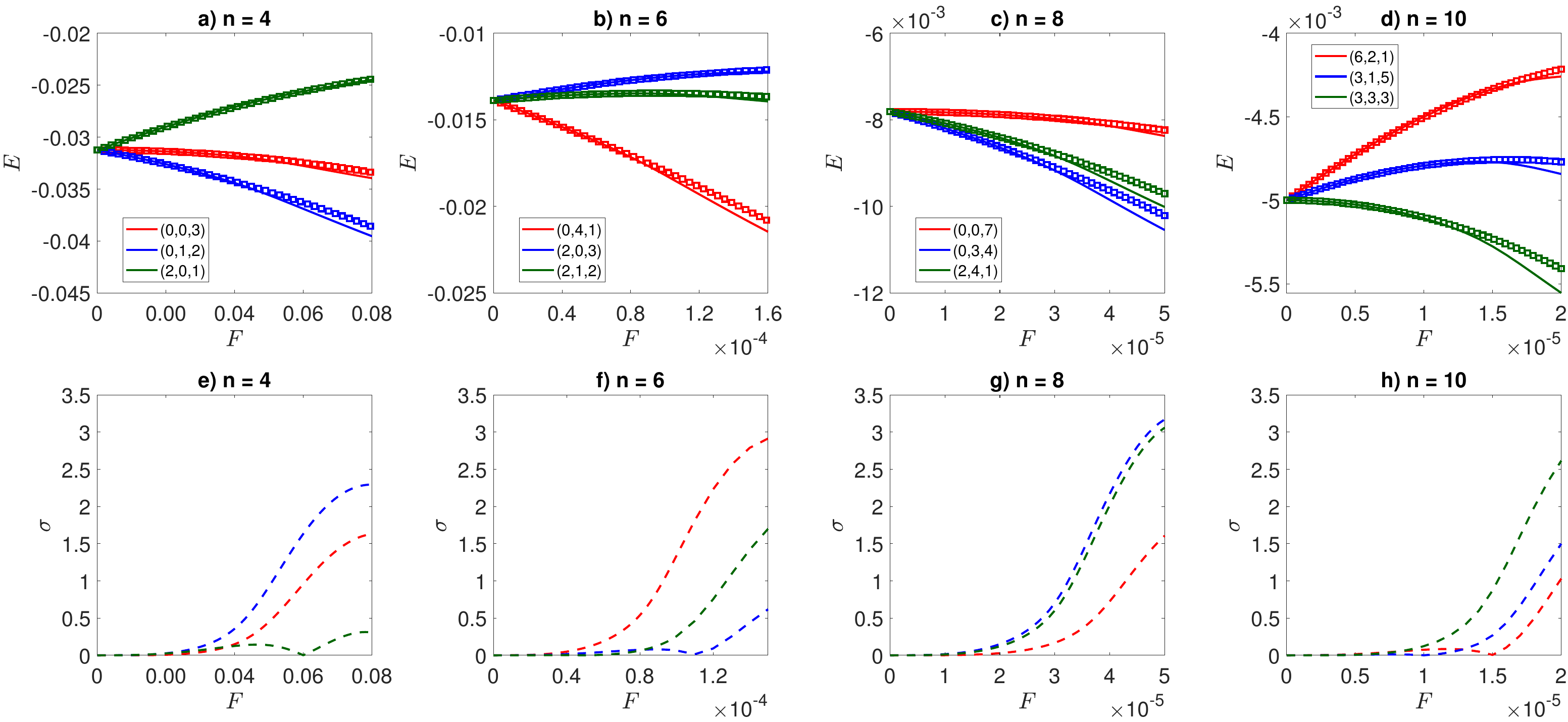}
	\caption{Same as Fig. \ref{fig:gs_excited} for a) $n=4$, b) $n=6$, c) $n=8$, and d) $n=10$.} 
	\label{fig:excited_state}
\end{figure}

Figure \ref{fig:excited_state} displays the comparison for the excited states with the principle quantum number $n=\{4,6,8,10\}$.  In these cases, the relative deviations are less than $3.5\%$, which is still an acceptable value.

\section{Conclusions}
\label{conclusions}
In conclusion, we revisited the quadratic Stark effect of hydrogen. We rely on the time-independent non-degenerate perturbation theory and the general integral formula for associated Laguerre polynomials. We have also investigated the applicability interval of the approximated formula describing this effect using a numerical technique based on the Siegert State method. The comparison demonstrates that when the electric field is suitably weak, the approximated formula agrees well with numerical calculations not only for the ground state but also for highly excited states. The approximated formula can therefore be used to predict the eigenvalues of highly excited states in the presence of an exceedingly weak electric field for future high-precision numerical calculations. In addition, we emphasize that, in quantum mechanics, perturbation theory is not complicated but never trivial. To address the problem of a stronger electric field, it should be noted that higher-order corrections or improved approaches should be considered; however, these are not our objectives. In conclusion, we anticipate that the results of this study will serve as a valuable resource for undergraduates studying quantum mechanics and be readily accessible to them. 

\section*{Data availability}
Data generated or analyzed during this study are provided in full within the published article.

\section*{Competing interests}
The authors declare there are no competing interests.

\appendix
\label{appA}
\section{Derivation of the general formula for the integral of associated Laguerre polynomials}	
Following the procedure proposed in Ref. \cite{tai2021analytical} to deduce the general formula for integrals of Hermite polynomials, this section derives the general formula for the integral of the associated Laguerre polynomials, which is given by 
\begin{equation}
	\label{laguerreint} 
	Z(\alpha,k,k^\prime) = \int\limits_{0}^{\infty }x^{\alpha}u_{k,m}\left(x\right)u_{k^\prime,m}\left(x\right)dx,
\end{equation}
where 
\begin{equation}
	\label{udefinition} 
	u_{k,m}\left(x\right) = x^{m/2}\exp(-x/2)L_{k}^{m}(x),
\end{equation}
where $L_{k}^{m}(x)$ is the associated Laguerre polynomial and $\alpha$ is the non-negative integer, respectively. The two associated Laguerre polynomials in Eq. \eqref{laguerreint} are then expanded by the generating function \cite{weber2003essential} as shown:
\begin{align}
	\label{lague1}
	U(x,t)&=\sum\limits_{k=0}^{\infty }L_k^m(x)t^k=\dfrac{1}{(1-t)^{m+1}}\exp \left( -\dfrac{xt}{1-t} \right), \\
	\label{lague2}
	V(x,y)&=\sum\limits_{k^\prime =0}^{\infty }L_{k^\prime}^m(x)y^{k^\prime}=\dfrac{1}{(1-y)^{m+1}}\exp\left(-\dfrac{xy}{1-y}\right),
\end{align}
where $|t|<1$, and $|y|<1$. Multiplying Eqs. \eqref{lague1} and \eqref{lague2} side by side and by $x^{m+\alpha} \exp(-x)$, then taking the integral from 0 to $\infty$,  we get
\begin{multline}
	\label{expanding_equa} 
	\sum\limits_{k=0}^{\infty}\sum\limits_{k^\prime=0}^{\infty} \left[\int\limits_{0}^{\infty} x^{m+\alpha}\exp(-x) L_{k}^{m}(x)L_{k^\prime}^{m}(x)dx \right]  t^ky^{k^\prime} =  \\ \dfrac{1}{(1-t)^{m+1}(1-y)^{m+1}}\int\limits_{0}^{\infty }x^{m+\alpha}\exp \left[-x\left( 1+\dfrac{t}{1-t}+\dfrac{y}{1-y} \right) \right]dx. 
\end{multline}
The result of the integral on the right hand side (RHS) of Eq. \eqref{expanding_equa} can be straightforwardly obtained, 
\begin{equation}
	\label{rhsint} 
	I = \int\limits_{0}^{\infty }x^{m+\alpha}\exp\left[-x\left( 1+\dfrac{t}{1-t}+\dfrac{y}{1-y} \right) \right]dx 
	= \dfrac{(m+\alpha)!(1-t)^{m+\alpha+1}(1-y)^{m+\alpha +1}}{(1-ty)^{m+\alpha+1}},
\end{equation}
by applying the integral 
$	\int\limits_{0}^{\infty}x^n\exp(-px)dx=\dfrac{n!}{p^{n+1}}.
$ 
Substituting Eq. \eqref{rhsint} back into the RHS of Eq. \eqref{expanding_equa}, we obtain
\begin{equation}
	\label{newrhs}
	RHS = \dfrac{(m+\alpha)!(1-t)^{\alpha}(1-y)^{\alpha}}{(1-ty)^{m+\alpha+1}}.
\end{equation}
The binomials in Eq. \eqref{newrhs} are then expanded as
\begin{align}
	\label{a9} 
	(1-ty)^{-(m+\alpha+1)} & = \sum\limits_{s=0}^{\infty} { m+s+\alpha \choose s } (ty)^s,\\
	\label{a10} 
	(1-t)^\alpha(1-y)^\alpha & = \sum\limits_{i=0}^{\alpha} \sum\limits_{j=0}^{\alpha} (-1)^{i+j} {\alpha \choose i}{\alpha \choose j}t^iy^j.
\end{align}
Hence, the RHS can be rewritten as
\begin{equation}
	RHS = \sum_{s=0}^\infty \sum_{i=0}^\alpha \sum_{j=0}^\alpha (-1)^{i+j} {\alpha \choose i}{\alpha \choose j} { m+s+\alpha \choose s }t^{s+i}y^{s+j}. 
\end{equation}
The value of $Z(\alpha,k,k^\prime)$ is obviously the coefficient of $t^{s+i}y^{s+j}$ and can therefore be determined by unifying the coefficients of $t^{s+i}y^{s+j}$ from both sides of Eq. \eqref{expanding_equa}. We acquire
\begin{equation}\label{a13} 
	\begin{cases}
		s + i = k \\ 
		s + j = k^\prime
	\end{cases}
	\Rightarrow k^\prime = k - i + j.
\end{equation}	
Consequently, the desired formula for the general integral of the associated Laguerre polynomials is attained
\begin{equation}
	\label{laguerre_int} 
	Z(\alpha,k,k^\prime) = \sum\limits_{i=0}^{\alpha} \sum\limits_{j=0}^{\alpha} (-1)^{i+j} {\alpha \choose i}{\alpha \choose j} \dfrac{(m+\alpha+k-i)!}{(k-i)!}\delta_{k^\prime,k-i+j}.
\end{equation}
$Z(\alpha,k,k^\prime)$ is non-zero if, for a given set of $\alpha$ and $k$, only $k^\prime$ satisfies the selection rule determined by the Kronecker delta, $\delta_{k^\prime,k-i+j}$. To illustrate the application of the general formula Eq. \eqref{laguerre_int}, we take the known integral \cite{weber2003essential} 
\begin{equation}
	I_1 = \int\limits_0^\infty  u_{k,m}(x)u_{k^\prime,m}(x)dx = \int\limits_0^\infty x^m \exp(-x) L_k^m(x)L_{k^\prime}^m(x)dx= \dfrac{(k+m)!}{k!}\delta_{k^\prime,k},
\end{equation}
as an illustration. One can readily observe that the integral $I_1$ is equivalent to the case where $\alpha = 0$ and that the indices $i=j=0$. Substituting $\alpha=i=j=0$ back into Eq. \eqref{laguerre_int} yields
\begin{equation}
	Z(0,k,k^\prime) = (-1)^{0+0} {0 \choose 0}{0 \choose 0} \dfrac{(m+0+k-0)!}{(k-0)!}\delta_{k^\prime,k-0+0} = \dfrac{(k+m)!}{k!}\delta_{k^\prime,k}. 
\end{equation}
The well-known integral \cite{weber2003essential} 
\begin{equation}
	\label{I_2}
	I_2 = \int\limits_0^\infty x^{m+1} \exp(-x) L_k^m(x)L_{k^\prime}^m(x)dx = \dfrac{(k+m)!}{k!}(2k+m+1) \delta_{k^\prime,k},
\end{equation}
will be left as an exercise, and we encourage readers to derive the result in Eq. \eqref{I_2} on their own.

\section{The Siegert State method}	
In this section, we briefly present the Siegert State (SS) method that was used to numerically obtain the eigenvalues of Eq. \eqref{SE} for a given value $F>0$. Siegert \cite{siegert1939derivation} formulated a rigorous derivation of the Breit-Wigner resonance formula by solving the stationary Schr\"{o}dinger equation with the regularity and outgoing-wave boundary conditions. The SS eigenvalues given in the form of 
\begin{equation}
    \label{SS}
    E = \varepsilon - i \dfrac{\Gamma}{2}
\end{equation}
defines the energy $E$ and the ionization rate $\Gamma$ of the state. Here, $i$ denotes the imaginary unit. As been shown in Ref. \cite{batishchev2010atomic}, when the hydrogen atom is subject to a static electric field, its electron can be ionized even in a weak field by tunneling over the finite-size potential barrier to the classically allowed region. The real part of Eq. \eqref{SS}, $\varepsilon$, actually describes the Stark effect of hydrogen atom which is the main subject of this work.

In order to solve Eq. \eqref{SE} numerically by using the SS method, we write the ansatz for the solution of Eq. \eqref{SE} as 
\begin{equation}
    \label{ansatz}
    \psi(r) = f(\eta)\phi(\xi)\dfrac{e^{-im\varphi}}{\sqrt{2\pi}}. 
\end{equation}
Substituting Eq. \eqref{ansatz} into Eq. \eqref{SE} yields two coupled equations
\begin{align}
    \label{sss1}
    \left[\dfrac{d}{d \xi} \left(\xi \dfrac{d}{d \xi}\right) - \dfrac{m^2}{4\xi} + \dfrac{E\xi}{2} -\dfrac{F\xi^2}{4} \right]\phi(\xi) & = \beta_1(E)\phi(\xi) \\ 
    \label{sss2}
    \left[\dfrac{d}{d \eta} \left(\eta \dfrac{d}{d \eta}\right)  - \dfrac{m^2}{4\eta} + \dfrac{E\eta}{2} + \dfrac{F\eta^2}{4} \right]f(\eta)\ &= \beta_2(E)f(\eta),
\end{align}
where $\beta_1(E)$ and $\beta_2(E)$ depend on $E$ and must fulfill the constrain in the case of hydrogen atom
\begin{equation}
    \label{constrain}
    \beta_1(E) - \beta_2(E) = 1.
\end{equation}
To obtain the SS eigenvalues for given a set of quantum numbers $(n_1,n_2,m)$, following the approach in Ref. \cite{ohgoda2017photoionization}, we use the Newton-Raphson method to seek the root of the equation 
\begin{equation}
    g(E) = \beta_1(E) - \beta_2(E) - 1 = 0. 
\end{equation}
In practice, the bound-state energy when $F=0$, $E^{(0)} = -\dfrac{1}{2(n_1 + n_2 + m + 1)^2}$,  is chosen as the initial estimate for the Newton-Raphson method. Here, $\beta_1(E)$, $\beta_2(E)$ are determined by variationally solving Eqs. \eqref{sss1} and \eqref{sss2}. In doing so we expand the solutions to Eqs. \eqref{sss1} and \eqref{sss2} in the form 
\begin{align}
    \label{ssssss}
    \phi(\xi) &= \sum_{\mu=1}^{N} c_\mu  \xi^{m/2}\exp\left(-\xi/2\right)L_{\mu}^{m}\left(\xi\right) \\ 
    f(\eta) &= \sum_{\nu=1}^{N} c_\nu  \eta^{m/2}\exp\left(-\eta/2\right)L_{\nu}^{m}\left(\eta\right)
\end{align}
Subsequently, the matrix representation of Eqs. \eqref{sss1} and \eqref{sss2} can be constructed, and they can be diagonalized to produce their lowest eigenvalues, $\beta_1(E)$ and $\beta_2(E)$, respectively. It is worth mentioning that these matrices have the penta-diagonal band form and their elements can be determined by using the general formula for integral of associated Laguerre polynomials derived in Appendix A. 

In our numerical calculations, we have used $N = 300$ basis functions to expand $f(\eta)$ and $\phi(\xi)$, respectively and $\delta = 10^{-10}$ is the tolerance for the well-converged result in the Newton-Raphson method. For matrix diagonalization, we have employed the subroutine `zgeev' implemented in the LAPACK library.

\bibliography{ref.bib}

\end{document}